\documentclass[journal,10pt]{IEEEtran}
\IEEEoverridecommandlockouts
\usepackage{graphicx}
\usepackage{subcaption}  
\usepackage{multirow}
\usepackage{amsmath}
\usepackage{graphicx}
\usepackage{amssymb}
\usepackage{amsmath,amsthm} 
\usepackage{float}
\usepackage{enumerate}
\usepackage{epstopdf}
\usepackage{dblfloatfix}
\usepackage{caption}
\usepackage{amsfonts}
\usepackage{fancyhdr}
\usepackage{bbm}
\usepackage{cases}
\usepackage{cite}
\usepackage{breqn}
\usepackage{multirow}
\usepackage{mathrsfs}
\usepackage{algpascal}
\usepackage{algc}
\usepackage{algorithmicx}
\usepackage[ruled]{algorithm}
\usepackage{algpseudocode}
\usepackage{graphics}
\usepackage{epsfig}
\usepackage{mathrsfs} 
\usepackage{dsfont}
\usepackage{color}
\usepackage{xpatch}
\usepackage[hyphens]{url}
\usepackage{hyperref}

\floatname{algorithm}{Algorithm}

\algdef{SE}[DOWHILE]{Do}{doWhile}{\algorithmicdo}[1]{\algorithmicwhile\ #1}%

\ifCLASSINFOpdf
 \else
\fi

\begin{document}
\theoremstyle{definition} 
\newtheorem{theorem}{Theorem}[section]
\newtheorem{definition}[theorem]{Definition}
\newtheorem{lemma}[theorem]{Lemma}
\newtheorem{example}[theorem]{Example}
\newtheorem{Proposition}[theorem]{Proposition}
\newtheorem{Corollary}[theorem]{Corollary}

\title{Distributed Gossip-GAN for Low-overhead CSI Feedback Training in FDD mMIMO-OFDM Systems}

\author{
~\IEEEmembership{}Yuwen Cao\textsuperscript{},~\IEEEmembership{Member, IEEE},
Guijun Liu\textsuperscript{},
Tomoaki Ohtsuki\textsuperscript{},~\IEEEmembership{Senior Member, IEEE}, Howard H. Yang\textsuperscript{},~\IEEEmembership{Member, IEEE}, and Tony Q. S. Quek\textsuperscript{},~\IEEEmembership{Fellow, IEEE}

\thanks{
The work of Y. Cao was supported in part by the National Natural Science Foundation of China under Grant 62301143, in part by Shanghai Sailing Program under Grant 23YF1400800, and also in part by the Fundamental Research Funds for the Central Universities under Grant 2232024D-38. The work of T. Q. S. Quek was supported by the National Research Foundation, Singapore and Infocomm Media Development Authority under its Future Communications Research \& Development Programme.}
\thanks{Y. Cao and G. Liu are with the College of Information Science and Technology, Donghua University, Shanghai 201620, China. 
T. Ohtsuki is with the Department of Information and Computer Science, Keio University, Yokohama 223-8522, Japan. H. H. Yang is with Zhejiang University/University of Illinois at Urbana-Champaign Institute, Zhejiang University, Haining 314400, China. T. Q. S. Quek is with the Information Systems Technology and Design Pillar, Singapore University of Technology and Design, Singapore 487372.}
}

{}

\maketitle

\begin{abstract}
The deep autoencoder (DAE) framework has turned out to be efficient in reducing the channel state information (CSI) feedback overhead in massive multiple-input multiple-output (mMIMO) systems. 
However, these DAE approaches presented in prior works rely heavily on large-scale data collected through the base station (BS) for model training, thus rendering excessive bandwidth usage and data privacy issues, particularly for mMIMO systems. 
When considering users' mobility and encountering new channel environments, the existing CSI feedback models may often need to be retrained. Returning back to previous environments, however, will make these models perform poorly and face the risk of catastrophic forgetting.
To solve the above challenging problems, we propose a novel gossiping generative adversarial network (Gossip-GAN)-aided CSI feedback training framework. Notably, 
Gossip-GAN enables the CSI feedback training with low-overhead while preserving users' privacy.
Specially, each user collects a small amount of data to train a GAN model. Meanwhile, a fully distributed gossip-learning strategy is exploited to avoid model overfitting, and to accelerate the model training as well.  
Simulation results demonstrate that Gossip-GAN can i) achieve a similar CSI feedback accuracy as centralized training with real-world datasets, ii) address catastrophic forgetting challenges in mobile scenarios, 
and iii) greatly reduce the uplink bandwidth usage. 
Besides, our results show that the proposed approach possesses an inherent robustness. 
\end{abstract}


\section{Introduction}
Massive multiple-input multiple-output (mMIMO) systems deploy a large number of antennas at base stations (BSs), which can greatly enhance the performance in terms of the system capacity, spectrum efficiency, and data throughput rates \cite{ref1}. 
In addition, mMIMO is an
enabler for the future digital society infrastructure that will connect the Internet of people
and Internet of Things (IoTs) with network edge as well as other
network infrastructure \cite{ref2,mao1}.
As such, mMIMO has proven to be a crucial technology for the sixth-generation (6G) mobile communication system\cite{ref3,more1,more2,mao3,mao2}. However, these benefits of mMIMO systems can be realized only when the transmitter, especially the BS, has observed an accurate downlink channel state information (CSI)\cite{more3,more4}. For time division duplex (TDD) systems, the downlink channel can be obtained from the uplink channel through the channel reciprocity. However, for frequency division duplex (FDD) systems, there is no such channel reciprocity for the downlink channel. In this case, user equipments (UEs) shall first estimate the channel and then feed it back to the BS. Nevertheless, the feedback of large CSI in mMIMO-FDD systems in practice will consume a large amount of uplink transmission bandwidth \cite{ref4,alsabah20216g}. 

Traditional CSI feedback compression technologies such as i) compressed sensing (CS)\cite{gao2018compressive,choi2017compressed} and ii) codebook based schemes\cite{ref6,qin2023review}, are adopted in mMIMO systems to reduce the high feedback overhead in mMIMO-FDD systems. However, these methods are limited to either requiring a complex iterative process or using the CSI prior sparsity assumption. In future 6G mobile communication systems, these methods may be impractical. 
 
On the other hand, deep neural network (DNN) has been applied in various fields of wireless communications because of its excellent fitting ability. For CSI feedback compression, Wen \textit{et al.} \cite{ref7} propose a novel autoencoder-based CSI feedback framework (CsiNet) where the encoder in UE first compresses high dimensional CSI, and the decoder recovers compressed CSI accordingly. Compared with traditional algorithms, CsiNet shows excellent compression capability and reconstruction accuracy. Based on the CsiNet architecture, many new novel network architectures have been developed, e.g., CRNet\cite{ref8}, CsiNet-KD\cite{ref01}, TransNet\cite{ref9}, and DCRNet\cite{ref10}. Specifically, CRNet utilizes a novel multi-resolution architecture to improve the CSI feedback performance. In addition, CsiNet-KD incorporating knowledge distillation into CsiNet achieves the same compression performance with fewer parameters. In addition, TransNet and DCRNet use attention and dilated convolution, respectively, to replace traditional convolution operations, so as to obtain greater feature extraction capability and improve the CSI feedback performance.

\begin{figure}
    \centering
    \includegraphics[width=8cm,height=5cm]{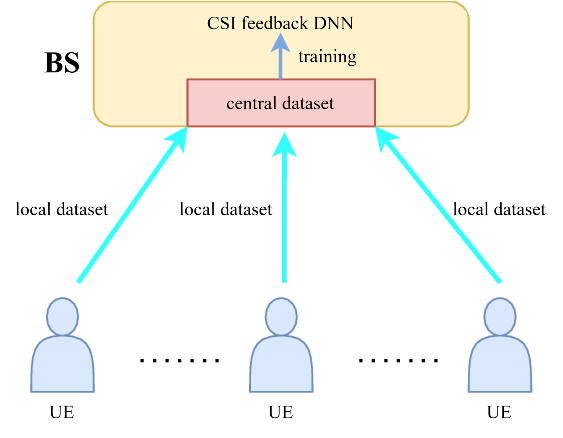}
    \caption{
    The central dataset training scenario for CSI feedback at the BS. 
    } 
    \label{fig:fig1}
\end{figure}

However, the above-mentioned DNN-based CSI feedback training frameworks assume that the DNN is trained using a central dataset collected at BS as shown in Fig. \ref{fig:fig1}. Moreover, delivering the local datasets to the BS will consume a large amount of bandwidth, particularly for the mMIMO systems. Besides, the data privacy problem will occur when encountering the uplink data leakage. In this context, the authors in \cite{ref11} claim that using generative model variational auto-encoder (VAE) can generate high-quality fake datasets to address potential privacy risks. The authors in \cite{ref12} utilize generative adversarial network (GAN) to model the channel distribution, and prove that the channels generated by GAN are consistent with the real channel distribution. However, training generative models such as the VAE and GAN usually require a large amount of data and impose excessive floating point operations per second (FLOPs)\cite{ref02}. 
Furthermore, when taking the fact into accounts, these UEs are usually resource-constrained as shown in Fig. \ref{fig:fig1}. Hence, the above-mentioned techniques may be impractical for resource-constrained scenarios. Therefore, it is believed that constructing an efficient generative model is crucial for achieving improved performance while preserving users’ privacy.

In addition, DNN typically lacks strong generalization capabilities. Considering the mobility of users, DNN-based solutions would need to recollect data and retrain in order to maintain high performance when users move to a new environment. To maintain high DNN training performance while tackling the challenges brought by the frequent user movements, transfer learning and a variety of online learning-based strategies have been proposed to enable a rapid adaptation to dynamic changing environments \cite{ref13,ref15,ref16,ref17}. However, these works usually ignore the catastrophic forgetting phenomenon that occurs when the user returns back to the prior channel environments. The lack of generalization of DNN involved in these works results in poor model performance when the user revisits the earlier environment. Consequently, this leads to the need for repeated data collection and model retraining\cite{ref14}. 

Based on the aforementioned 
discussions, the current existing CSI feedback training still suffers from the following challenges:

\begin{itemize}
\item Most works assume that CSI feedback training is done at the BS. The dataset collection at the BS will cost huge uplink transmission bandwith and lead to potential data privacy problems. Thus, it is critical to explore new low-overhead CSI feedback training framework.
\item The training of generative models necessitates substantial data and computational power, thereby leading to extended data collection and model training duration. In CSI feedback scenarios, UEs typically possess limited hardware resources, which necessitates cost-effective and rapid model training technologies.
\item The practical wireless channel distribution is complex. To capture the channel distribution and produce high-quality channel data, an appropriate generative model needs to be studied and employed.
\item In most prior training methods, when users move to a new environment, CSI feedback DNN needs to keep adapting to the new environment while avoiding forgetting the past scenes. This indicates that new techniques that can help to improve the generalization capabilities of neural networks in CSI feedback training deserve a further study. 
\end{itemize}

In this work, we propose a novel Gossip-GAN based CSI feedback training framework for the mMIMO systems. We design a lightweight gossip learning (GL) strategy for multi-users' cooperation.
The Gossip-GAN framework is fully distributed and the GAN model is designed based on the consistency regularization GAN (CTGAN) strategy \cite{ref22}. 
This means that the Gossip-GAN framework has the potential to provide high-quality channel modeling capabilities while incurring low uplink transmission overhead. Notably, unlike federated learning (FL)\cite{hegedHus2021decentralized}, which relies on communication between a central server and distributed local clients, GL operates as a fully decentralized distributed-learning framework\cite{ormandi2013gossip,ref16}. It avoids periodic information exchange between central server (i.e., BS) and local clients (i.e., UEs) and is valid to reduce uplink communication consumption. The details of the Gossip-GAN framework will be introduced later. 
Besides, we further explore the Gossip-GAN framework to collaborate with CSI feedback DNN to adapt to new environments while avoiding catastrophic forgetting problems. 

In summary, the main contributions of this work are summarized as follows:
\begin{itemize}
\item This paper proposes a novel Gossip-GAN-based CSI feedback training framework that needs low-overhead uplink transmission and avoids the potential UE data privacy problems. The feedback performance of our framework is comparable to that of previous methods based on centralized training, which require significant transmission overhead, and can even be slightly improved.
\item By combining our distributed GL strategy and the state-of-the-art generation model, we can train a generation model (GAN) with excellent performance in the case of resource-limited users in CSI feedback scenarios. Meanwhile, compared to existing central generative model training methods applied in wireless communication, the computation cost can be greatly reduced.
\item To solve the catastrophic forgetting problem and improve the generalization ability of the model, we propose to use the GAN model obtained by the proposed framework to assist in training when facing new scenarios. Compared to the previous method \cite{bibcl}, which is based on storing data, this method enables better generalization performance and less storage space. To the best of our knowledge, this is the first work that uses the generative model to solve catastrophic forgetting problems for CSI feedback.
\item In particular, most of the works like \cite{ref8,ref10,ref11} use the COST2100 dataset\cite{ref7,cost2100} to evaluate the model performance. However, practical channels are often more complex, and we use a more realistic 3D ray tracing DeepMIMO dataset\cite{deepmimo} to verify the effectiveness of our proposed framework.
\item The simulation results show that our approach can achieve similar performance as training by recollecting data. 
\end{itemize}
  \begin{figure*}[t!]
    \centering
    \includegraphics[width=16cm]{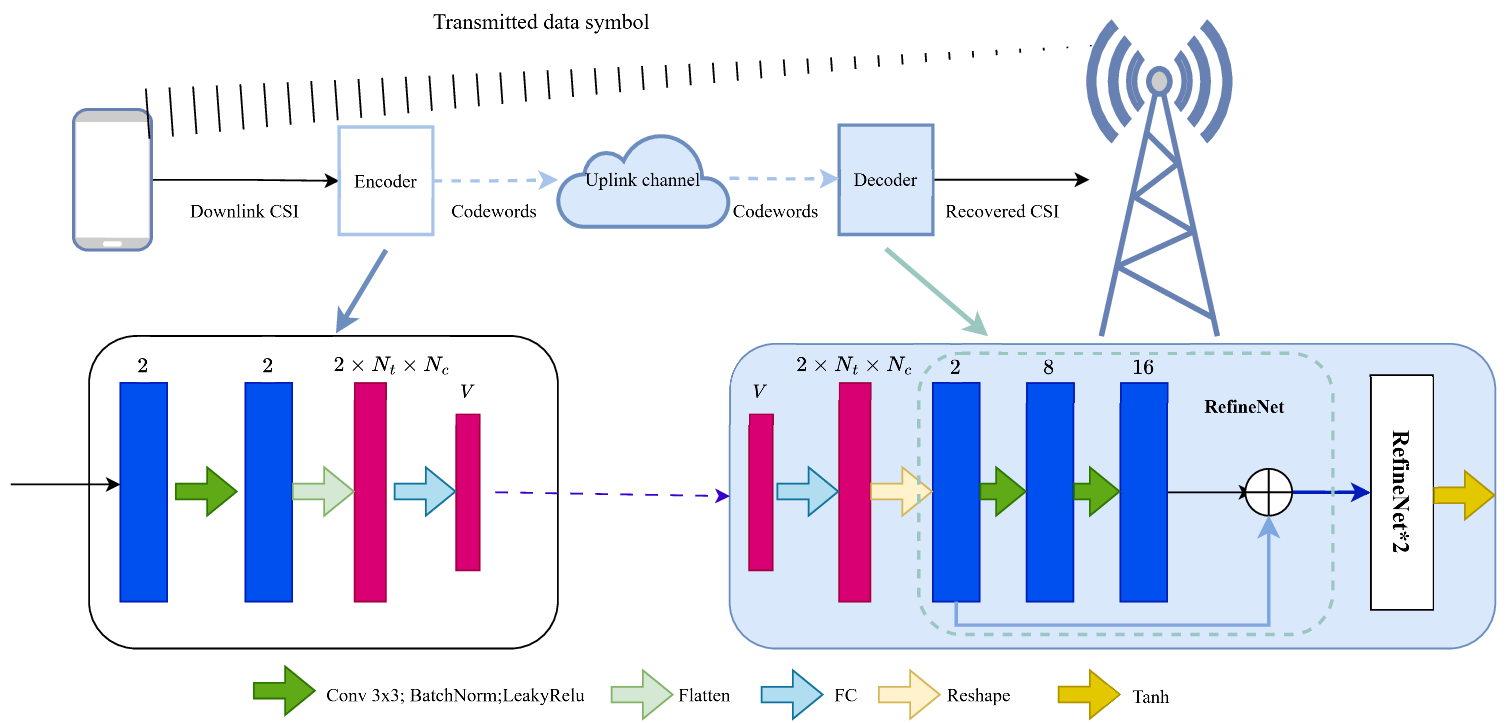}
    \caption{
    An overview of the autoencoder CSI feedback framework and the basic neural network architecture adopted in this paper. 
    } 
    \label{fig:fig2}
\end{figure*} 
The rest of this paper is organized as follows. Section \uppercase\expandafter{\romannumeral2} introduces the system model and the deep autoencoder (DAE) framework for CSI feedback. Section \uppercase\expandafter{\romannumeral3} presents our proposed novel Gossip-GAN framework. Section \uppercase\expandafter{\romannumeral3} provides the simulation results. Section \uppercase\expandafter{\romannumeral4} concludes our paper and shows some future research directions.

{\it Notations}: 
The lowercase letters represent scalars. 
Bold uppercase and lowercase letters are used to
represent matrices and vectors, respectively. 
In addition, $\left( {\cdot} \right)^{T}$, $\left( {\cdot} \right)^{H}$, and $\left\| {\cdot} \right\|_{\rm{2}}$ denote the transpose, complex conjugate, and $l_{2}$-norm operations of a matrix, respectively. $\mathbb{C}$ denotes the complex space. $\mathbb{E}\left[\cdot\right]$ means the expectation of the argument and \( \otimes \) represents the Kronecker product. 

\section{System Model}
\subsection{The FDD mMIMO-OFDM System}

Consider an FDD mMIMO system, where the BS is equipped with
$N_{t}$ transmit antennas with $N_{t} \gg 1$ and serves multiple single-antenna UEs within its coverage. Orthogonal frequency division multiplexing (OFDM) is used over $N_{c}$ subcarriers for downlink transmission.
%
Let $x_{n}$ denote the transmitted data symbol over the $n$-th subcarrier. For the $n$-th subcarrier, the received signal $y_{n}$ can be represented as
\begin{equation} \label{equ:1}
y_{n}=\mathbf{h}_{n}^H\mathbf{v}_{n}x_{n}+z_{n},
\end{equation}
where $\mathbf{h}_n\in\mathbb{C}^{N_t\times1}$ corresponds to the channel vector, $\mathbf{v}_{n}$ denotes the precoding vector, and $z_{n}$ is the additive complex noise. 

The channel vector $\mathbf{h}_n$ of the FDD mMIMO-OFDM system consists of $L$ paths. In particular, we consider an outdoor channel model whose 3D ray-tracing parameters are generated via \textit{DeepMIMO}.\footnote{More details regarding the channel distribution information and the dataset generation codes are available on https://www.deepmimo.net/.} 
We denote the receive power by $ \rho_{l}$, the phase by $ \vartheta_{l} $, and the propagation delay by $ \tau_{l} $ in the channel path $l$ between the BS and the UE. In addition, let $B$ denote the system bandwidth and $d$ be the antenna spacing.
As such, the channel on sub-carrier $n$ can be modeled as\cite{deepmimo}:
\begin{equation} \label{equ:2}
\mathbf{h}_n=\sum_{l=1}^L\sqrt{\frac{\rho_{l}}{N_c}}e^{j\left(\vartheta_{l}+\frac{2\pi n}{N_c}\tau_{l}B\right)}\mathbf{a}(\phi_{az},\phi_{el}), 
\end{equation}
where $\mathbf{a}(\phi_{az},\phi_{el})$ is the array response vector of the BS, 
and can be expressed as:
\begin{equation}
\mathbf{a}\left(\phi_{az}, \phi_{el}\right) = \mathbf{a}_z\left(\phi_{el}\right) \otimes \mathbf{a}_y\left(\phi_{az}, \phi_{el}\right) \otimes \mathbf{a}_x\left(\phi_{az}, \phi_{el}\right),
\end{equation}
where $\phi_{az}$ and $\phi_{el}$ represent the azimuth and elevation angles of departure from the BS,  respectively. In addition,
\( \mathbf{a}_x(\cdot) \), \( \mathbf{a}_y(\cdot) \), and \( \mathbf{a}_z(\cdot) \) denote the BS array response vectors in the \( x \), \( y \), and \( z \) directions, respectively, and are expressed as:
\begin{equation}
        \begin{aligned}
\mathbf{a_x}\left(\phi_{\mathrm{az}},\phi_{\mathrm{el}}\right) &= \left[1,e^{jnd\sin(\phi_{\mathrm{el}})\cos(\phi_{\mathrm{az}})},\ldots\right. \\
    &\left.\ldots,e^{jnd(N_{t}-1)\sin(\phi_{\mathrm{el}})\cos(\phi_{\mathrm{az}})}\right]^{T},
    \end{aligned}
\end{equation}
\begin{equation}
        \begin{aligned}
\mathbf{a_y}\left(\phi_{\mathrm{az}},\phi_{\mathrm{el}}\right) &= \left[1,e^{jnd\sin(\phi_{\mathrm{el}})\sin(\phi_{\mathrm{az}})},\ldots\right. \\
    &\left.\ldots,e^{jnd(N_{t}-1)\sin(\phi_{\mathrm{el}})\sin(\phi_{\mathrm{az}})}\right]^{T},
    \end{aligned}
\end{equation}
\begin{equation}
\mathbf{a}_z\left(\phi_{el}\right) = 
\begin{bmatrix}
1, e^{jnd\cos\left(\phi_{el}\right)}, \dots, e^{jnd(N_t-1)\cos\left(\phi_{el}\right)}
\end{bmatrix}^T.
\end{equation}

The entire channel matrix estimated by UE can be denoted as
$
\mathbf{H} = [\mathbf{h}_1, \mathbf{h}_2, \ldots, \mathbf{h}_{N_c}] \in \mathbb{C}^{N_t \times N_c}.
$ In mMIMO systems, the channel matrix is of high dimension, which should be compressed and then fed back to the BS to avoid large transmission overhead.

\subsection{Deep Autoencoder CSI Feedback Framework}
To reduce the CSI feedback overhead, the DAE is adopted to compress CSI and then recover it. More concretely, UEs first convert $\mathbf{H}$ into a real matrix of size $N_t \times N_c \times 2$ and then compress the matrix into low-dimensional codewords using the encoder network. Afterward, UEs feed codewords back to the BS according to the following encoding criterion:
\begin{equation} \label{equ:3}
\mathbf{s}=f_{enc}(\mathbf{H}),
\end{equation}
where $\mathbf{s} \in \mathbb{C}^{V \times 1}$ denotes the compressed codewords, $V$ represents the size of the codewords, 
and $f_{enc}(\cdot)$ refers to the encoder neural network.

After receiving the codewords $\mathbf{s}$ from UEs, the BS utilizes the decoder network to recover the original channel matrix $\mathbf{H}$ based on the following decoding criterion:
\begin{equation} \label{equ:4}
\hat{\mathbf{H}}=f_{dec}(\mathbf{s}),
\end{equation}
where $
\hat{\mathbf{H}} \in \mathbb{C}^{N_t \times N_c}
$ denotes the CSI matrix and 
$f_{dec}(\cdot)$ represents the decoder neural network. The DNN's strong fitting ability yields a very high precision of the restored $\hat{\mathbf{H}}$.

The framework overview and the adopted autoencoder structure are illustrated in Fig. \ref{fig:fig2}.\footnote{Note that the proposed method is compatible with the other state-of-the-art CSI feedback training networks. For simplicity, we use the CsiNet structure in this work. In Section IV, we demonstrate the compatibility of our method with the other state-of-the-art architectures.} 
The encoder part uses 3$\times$3 convolutional kernels and batch-normal layers\cite{ref7} to extract features. It then employs a fully connected (FC) layer to compress the CSI into low-dimensional codewords, which are then fed back to the BS. In this case, the compressed ratio denoted by $\gamma$ can be defined as 
\begin{equation} \label{equ:5}
\gamma =\frac{V}{2N_{t}N_{c}}. 
\end{equation}
The decoder uses an FC layer to restore the codewords back to the original CSI size. Then, it employs three RefineNet structures to enhance the model's fitting capability, and finally recovers the full CSI\cite{ref30}. For the last layer, we use the tanh\cite{dubey2022activation} activation function, while for all other layers, we use leakyReLU. We adopt the mean square error (MSE) as the loss function, which is given by
\begin{equation} \label{equ:6}
Loss=\frac{1}{N_{s}}\sum_{i=1}^{N_s}||\mathbf{H}_{i}-\hat{\mathbf{H}}_{i}||_{2}^{2}, 
\end{equation}
where $ N_{s} $ indicates the total number of data samples.
\( \mathbf{H}_i \) and \( \hat{\mathbf{H}}_i \) represent the \( i \)th data and recovered CSI data, respectively.\footnote{The challenging research problems considered in our FDD mMIMO-OFDM system are highlighted below. Note that in mMIMO systems, the channel matrix is of high dimension, which will cause huge uplink transmission bandwidth when gathering and feeding back the dataset to the BS. In addition, data privacy risks may be encountered during the dataset collection process at the BS. To overcome this challenge, conventional CSI feedback training models compressing the CSI matrix and then feeding back to the BS have been developed. However, previous works \cite{ref7}, \cite{ref8} train the generative models using substantial data and computational power, thus leading to extended data collection and model training overhead. More importantly, when UE moves to a new environment, the current existing CSI feedback DNN training models are invalid to adapt to new channel environment.}

It is noted that, compared with the previous centralized learning, as well as the prior decentralized distributed-learning frameworks \cite{ormandi2013gossip, ref21,hegedHus2021decentralized,ref16}, our Gossip-GAN framework can provide the following important properties: 
\begin{itemize}
    \item To reduce resource consumption while maintaining good learning performance, in our framework, the BS picks a subset of $K$ UEs within its coverage to 
train their local GAN generators, and then forward one trained generator to the BS to
assist the CSI feedback model training with low transmission overhead.
\item In our framework, only a small amount of data is collected from each of the selected $K$ UEs, and the system uploads only a minimal number of neural network parameters. Thus, our framework will greatly relieve burdens on existing communication systems, particularly for those systems with resource-constrained devices, e.g., IoT devices, and mobile phones.
\item The well-trained CSI feedback model  in return can be used as a general model which is applicable to other users distributed in this region.
\end{itemize}

\section{Proposed Gossip-GAN Training Framework}
In this section, we first introduce the GAN model we used. Then, we present the low-overhead GL GAN strategy we designed and its advantages. Finally, we introduce Gossip-GAN for solving the challenging catastrophic forgetting problem.

\subsection{The Adopted GAN Model}
In order to address the risk of data leakage and user privacy issues, we can model the true distribution of the channel: $P_{r}(\mathbf{H})$, and then sample data from $P_{r}(\mathbf{H})$ to perform model training.\footnote{Note that, we ignore the adversarial machine learning (ML) threats in the CSI feedback in our model. A similar assumption has also been made in references \cite{ref12}, \cite{ref13}.} However, due to the complexity of the channel environment, traditional channel modeling methods find it difficult to accurately model $P_{r}(\mathbf{H})$. 
For instance, geometry-based stochastic channel models (e.g., 3GPP TR 38.901 \cite{docomo20165g}) exhibit known limitations due to the presence of strong reflections and scattering, as well as their heavy reliance on predefined statistical assumptions and geometry parameters \cite{docomo20165g}. As a generative model, GAN has shown powerful modeling capabilities\cite{ref12} for wireless communications. 

\begin{figure}
    \centering
    \includegraphics[width=9cm]{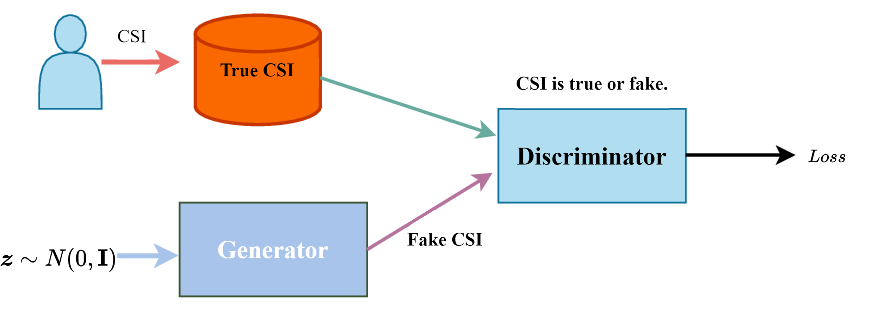}
    \caption{
    The working diagram of GAN model.
    } 
    \label{fig:fig3}
\end{figure}

As shown in Fig. \ref{fig:fig3}, GAN consists of a generator and a discriminator. The generator inputs normal distribution data and generates fake CSI data based on the channel distribution
$P_{f}(\tilde{\mathbf{H}})$ that it captures. The discriminator is used to determine whether the input CSI is real data collected by UEs or fake data generated by the generator. The $Loss$ for GAN training can be understood as the discriminator's accuracy in determining whether the CSI is true or fake. Using alternating training, the generator tries to generate data similar to true CSI: maximize the $Loss$, while the discriminator tries to judge the truth of the data: minimize the $Loss$. This adversarial approach will eventually allow the generator to model the real CSI distribution $P_{r}(\mathbf{H})$.

To speed up the convergence and guarantee accurate modeling ability during model training, we do not use classification loss such as cross entropy\cite{mao2023cross}. Instead, we adopt the earth mover (EM) distance as the loss function. The EM distance $Loss$ can be expressed as:
\begin{equation} \label{equ:7}
Loss(\mathbf{H},\tilde{\mathbf{H}})=\sup_{f\in\text{1-Lip}}\mathbb{E}_{\mathbf{H}\sim P_r(\mathbf{H})}[f(\mathbf{H})]-\mathbb{E}_{\tilde{\mathbf{H}}\sim P_{f}(\tilde{\mathbf{H}})}[f(\tilde{\mathbf{H}})], 
\end{equation}
where $\text{1-Lip}$ denotes the 1-Lispschitz function\cite{gulrajani2017improved} and $f(\cdot)$ denotes the discriminator network. 
Compared to the direct classification loss, the EM distance adds the 1-Lipschitz function constraint to the model, making the model training process more convergent when the difference between the distributions of $P_r(\mathbf{H})$ and $P_{f}(\tilde{\mathbf{H}})$  is significant
and therefore results in superior performance\cite{ref22}. 

Let $G(\cdot)$ and $D(\cdot)$ denote the generator and discriminator, respectively. Equation (\ref{equ:7}) can be implemented as (\ref{equ:ge}) for the generator and (\ref{equ:8}) for the discriminator, as proposed by CTGAN \cite{ref22}. Namely, we have
\begin{equation} \label{equ:ge}
Loss(\boldsymbol{z})=-D(G(\boldsymbol{z})),
\end{equation}
\begin{equation} \label{equ:8}
\begin{aligned}
Loss(\mathbf{H},\tilde{\mathbf{H}})=& \mathbb{E}_{\boldsymbol{z}\sim N(0,\mathbf{I})}\left[D(G(\boldsymbol{z}))\right]-\mathbb{E}_{\mathbf{H}\sim P_r(\mathbf{H})}[D(\mathbf{H})] \\
& +\lambda_{1}\mathbb{E}_{\widehat{\mathbf{H}}}[\Big(\|\nabla_{\widehat{\mathbf{H}}}D(\widehat{\mathbf{H}})\|_{2}-1\Big)^{2}]+ \\
&\lambda_2\mathbb{E}_{\mathbf{H}\thicksim P_r(\mathbf{H})}[\max(0,\|D_{1}(\mathbf{H})-D_{2}(\mathbf{H}) \|_2+ \\
&0.1\cdot \|D_{1}^{\prime}(\mathbf{H})-D_{2}^{\prime}(\mathbf{H})\|_2-M^{\prime})], 
\end{aligned}
\end{equation}
where $\widehat{\mathbf{H}}=i\mathbf{H}+(1-i)G(\mathbf{z})$, 
with $i$ following a uniform distribution.   
$\nabla_{\widehat{\mathbf{H}}}D(\widehat{\mathbf{H}})$ 
denotes the gradient of $\widehat{\mathbf{H}}$. 
$D_{1}(\mathbf{H})$ and $D_{2}(\mathbf{H})$ denote the outputs of the discriminator when two dropout probabilities are applied to $D(\cdot)$ when $\mathbf{H}$ is input. $D_{1}^{\prime}(\mathbf{H})$ and $D_{2}^{\prime}(\mathbf{H})$ represent the output of the second-to-last layer of $D(\cdot)$ with dropout layers. 
$\lambda_1$, $\lambda_2$, and $M^{\prime}$ are hyperparameters. 

\begin{table}[t!]
\centering
\caption{The Structure of Generator $G(\cdot)$.}  
\begin{tabular}{|l|p{4.5cm}|}
\hline
\textbf{Layer} & \textbf{Details} \\
\hline
Input & Input $\mathbf{z}$ of size: $128\times1\times1$. \\
\hline

\hline
ConvTranspose2d\cite{pytorch} & Transposed convolution layer with 128 input channels, 64 output channels, and a kernel size of $4\times4$. \\
\hline
ResidualBlock$\times3$: upsample & 
Each ResidualBlock contains two ConvBlock(Conv2d + BatchNorm2d + ReLU) 
with input/output channels 64,  a kernel size of $3\times3$. 
The output from the first ConvBlock undergoes nearest-neighbor interpolation for upsampling with a scale factor of 2, and the upsampled result is then fed into the second ConvBlock.
Finally, the output of the second ConvBlock is added with the input that has undergone nearest-neighbor interpolation upsampling\cite{pytorch} with a scale factor of 2.
\\
\hline
BatchNorm2d & Batch normalization layer. \\
\hline
ReLU & ReLU activation function. \\
\hline
Conv2d & Convolutional layer with input channels 64, output channels 2, kernel size $3\times3$, padding 1. \\
\hline
Tanh & Tanh activation function. \\
\hline
\end{tabular}

\label{tab:generator_structure}
\end{table}

\begin{table}[ht]
\centering
\caption{The Structure of Discriminator $D(\cdot)$.}
\begin{tabular}{|l|p{4.5cm}|}
\hline
\textbf{Layer} & \textbf{Details} \\

\hline
Input & Input $\tilde{\mathbf{H}}$ of size: $2\times32\times32$. \\
\hline

ResidualBlock$\times2$: downsample &  
Each ResidualBlock contains two ConvBlock(Conv2d + BatchNorm2d + ReLU)
with output channels 64,  a kernel size of $3\times3$. 
The output from the first ConvBlock undergoes an avg-pooling layer\cite{avgpool} with kernel size $2\times2$ and stride 2. The downsampled result is then fed into the second ConvBlock.
Finally, the output of the second ConvBlock is added with the input that has passed through another avg-pooling layer with kernel size of $2\times2$ and stride 2.\\
\hline

Dropout & Dropout layer with probability $dropout$, applied after ResidualBlock. \\
\hline
ResidualBlock & This ResidualBlock has no avg-pooling layer, other structures are exactly the same as the ResidualBlock:downsample above.  \\
\hline
Dropout & Dropout layer with probability $dropout$, applied after ResidualBlock. \\
\hline
ResidualBlock & ResidualBlock with input/output channels 64, has the same structure as the previous ResidualBlock. \\
\hline
Dropout & Dropout layer with probability $dropout$, applied after ResidualBlock. \\
\hline
Mean layer & By averaging the 3rd and 4th dimensions of input, a 64-size feature vector is obtained. \\
\hline
Linear & Fully connected layer (Linear) with input size 64, output size 1. \\
\hline
\end{tabular}

\label{tab:discriminator_structure}
\end{table}

\begin{figure*}
    \centering
16    \includegraphics[width=18cm]{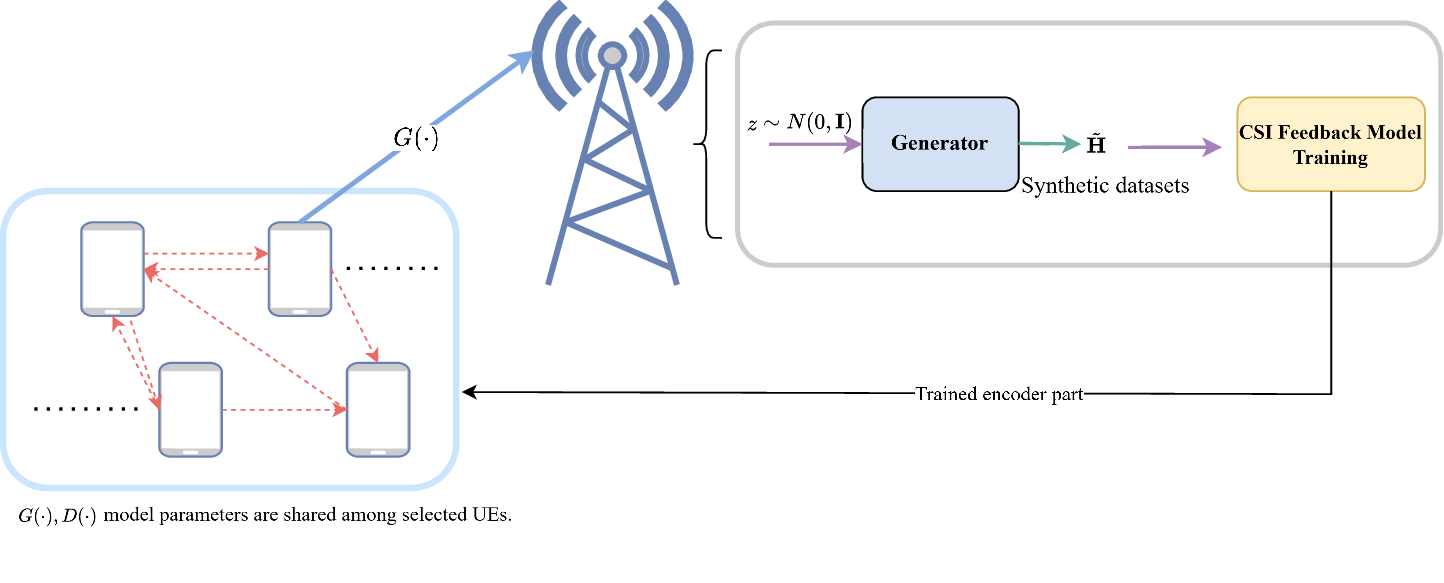}
    \caption{
The workflow diagram of the proposed Gossip-GAN CSI training framework, in which each UE trains a GAN model using a small amount of local dataset collected.} 
    \label{fig:fig4}
\end{figure*}

In our experiments, it is found that using $D_{1}^{\prime}(\mathbf{H}),D_{2}^{\prime}(\mathbf{H})$ can lead to only a slight performance improvement, and we will show the performance comparison between these in Section IV. In order to reduce computation, we finally adopt (\ref{equ:ge}) and (\ref{equ:9})  as loss function:
\begin{equation} \label{equ:9}
\begin{aligned}
Loss(\mathbf{H},\tilde{\mathbf{H}})&=\mathbb{E}_{\boldsymbol{z}\sim N(0,\mathbf{I})}\left[D(G(\boldsymbol{z}))\right]-\mathbb{E}_{\mathbf{H}\sim P_r(\mathbf{H})}[D(\mathbf{H})]
\\+&\lambda_1\mathbb{E}_{\widehat{\mathbf{H}}}[\left(\|\nabla_{\widehat{\mathbf{H}}}D(\widehat{\mathbf{H}})\|_2-1\right)^2]+\\
&\lambda_2\mathbb{E}_{\mathbf{H}\sim P_{r}(\mathbf{H})}[\max\left(0, \|D_{1}(\mathbf{H})-D_{2}(\mathbf{H}) \|_2 -M^{\prime}\right)
].
\end{aligned}
\end{equation}

Note that the structure of the GAN is detailed in TABLE \ref{tab:generator_structure} and TABLE \ref{tab:discriminator_structure}. If the stride is not mentioned, it is assumed to be 1 by default.

\subsection{Proposed Gossip-GAN CSI Feedback Training Framework}






\begin{algorithm}[t!]
\caption{GL Skeleton}
\label{alg1}
\begin{algorithmic}[1]
\State Collect local CSI measurements $\mathbf{H}$ at UE $i$
\State Initialize generator and discriminator: $G_i(\cdot), D_i(\cdot) \leftarrow \text{initialModel()}$
\State Perform local training: \textbf{LOCALTRAINING}($G_i(\cdot), D_i(\cdot)$)

\Function{LOCALTRAINING}{$G_i(\cdot),D_i(\cdot)$}

\While{stopping criterion not met}
    \State \textbf{Wait} for a fixed interval $\Delta$
    \State \textit{Select UE j} 
    \State Transmit models: $\textit{SEND}_{i \rightarrow j}(G_i(\cdot),D_i(\cdot),j)$
    \State On receiving model:\textbf{ONRECEIVEMODEL()}
\EndWhile

\EndFunction

\Function{ONRECEIVEMODEL}{}
    \State Save the received models: $\textbf{save}(G_j(\cdot), D_j(\cdot))$
    \If{Number of received  models $\geq n_{\text{peers}}$}
    \State $G_i(\cdot),D_i(\cdot) \leftarrow \textit{MERGE\_SAVED\_MODELS()}$
    \EndIf
\EndFunction
\end{algorithmic}
\end{algorithm}

The GAN training process typically requires a large number of CSI measurements and high FLOPs \cite{GANdrawback}. For the CSI feedback scenario, the UE usually has limited computing and storage resources, and can only collect a small amount of CSI data. However, the GAN model trained with a small amount of data does not perform well. For the case where each UE only collects a small amount of data, we use GL to implement distributed GAN training.

As shown in Fig. \ref{fig:fig4} and \textbf{Algorithm \ref{alg1}}, 
the GL uses a device-to-device (D2D) approach for training, without the involvement of the central server (i.e., the BS), thereby saving uplink bandwidth resources\cite{onoszko2021decentralized}. 
Each UE performs local GAN model training and
communicates with one another after a specified training time, sharing the parameters of both the generator and discriminator with selected peers.
Meanwhile, when the UE collects the model parameters from $n_{peers}$ other users, they perform a model merging operation. The merged model is then further trained locally.


\begin{figure}[!t]
    \centering
    \begin{subfigure}[t]{0.20\textwidth}
        \centering
        \includegraphics[width=\textwidth]{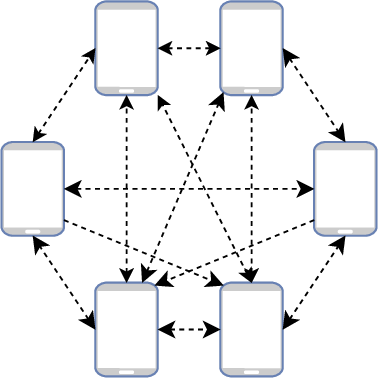} 
        \caption{ Topology 1.}
        \label{fig:topology1}
    \end{subfigure}
    \hspace{0.02\textwidth} 
    \begin{subfigure}[t]{0.20\textwidth}
        \centering
        \includegraphics[width=\textwidth]{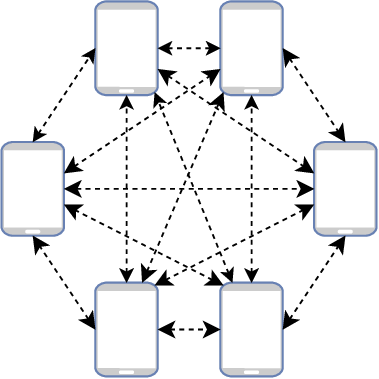} 
        \caption{Topology 2.}
        \label{fig:topology2}
    \end{subfigure}
    
    \caption{The diagrams of our proposed two different network topologies.
    In the first topology, as shown in Fig. 5(a), each UE randomly selects four other UEs to perform D2D communications simultaneously. In the second topology, as shown in Fig. 5(b), each UE communicates with all the other UEs simultaneously. Herein, it is noted that we let $K$ = 6 in this figure.}
    \label{fig:fig5}
\end{figure}

In \textbf{Algorithm 1}, the policy for selecting the UE and the number of $n_{peers}$ depends on the network connectivity architecture (i.e., topology)\cite{ref16}. In the following, we design two network topologies for Gossip-GAN training, as shown in Fig. \ref{fig:fig5}.\footnote{In our designed network connectivity Topology 1 for Gossip-GAN training, we assume that each participant of the GL framework will only contact with limited number of UEs due to the communication limitation.} The model parameters of $G$ and $D$ are transmitted in a fully distributed manner between the UEs.
A simple but effective model merging strategy is adopted as follows\cite{hardy2018gossiping}:
\begin{align}
\label{equ:10}
G_{i}(\cdot)=\frac{1}{n_{peers}}\sum_{j=1}^{n_{peers}}G_{j}(\cdot), 
\end{align}
\begin{align}
\label{equ:1002}
D_{i}(\cdot)=\frac{1}{n_{peers}}\sum_{j=1}^{n_{peers}}D_{j}(\cdot).
\end{align}

Specifically, each UE performs communication operations with its neighbors based on a fixed topology and at predefined time intervals. Once a UE receives model parameters from $n_{{peers}}$ peers, it proceeds with the model aggregation operation. Training then continues based on the aggregated model. 
The use of fixed communication intervals and  the average-based aggregation ensures a certain level of model consistency and synchronization across UEs, even in the absence of a central server.

During training, UE integrates the parameters of other UE models using the GL method, enabling the local GAN model to learn characteristics of other UEs' data. 
Each UE uses GL to train the model instead of using large amounts of data.
Clearly, adopting the distributed approach reduces UE's FLOPs and this GL parallel training mode accelerates the training speed.
Experimental results demonstrate that using the second topology structure can even approximate the performance of models trained on a central dataset. 

As shown in Fig. \ref{fig:fig4}, 
a randomly selected UE GAN generator trained via GL is transmitted to BS once it has been trained. BS uses the generator to produce a certain quantity of fake datasets, which are then employed for training a DAE model. Once the training for the autoencoder model is complete, BS transmits the decoder part to UEs.
The specified algorithm of the proposed Gossip-GAN is shown in \textbf{Algorithm \ref{alg2}}. 

\begin{algorithm}[t!]
\caption{The Proposed Gossip-GAN For CSI Feedback Training
}\label{alg2}
\begin{algorithmic}[1]
\For{$i = 1$ to $K$}  
\State Conduct local training GAN using \textbf{Algorithm 1}.
\EndFor
\State Randomly select $G(\cdot)$ from $\{G_1, G_2, \dots, G_K\}$ and transmit it to the BS. 
\State The BS generates a synthetic dataset $\mathcal{D}_{\text{gen}}$ defined as:
\begin{equation}
\mathcal{D}_{\text{gen}} = \{G(\mathbf{z}_j) \mid \mathbf{z}_j \sim N(0,\mathbf{I}), \; j=1, \dots, S\},
\end{equation}
where $S$ is the number of synthetic data samples.
\State Train the DAE-based CSI feedback framework by minimizing:
\begin{equation}
\min
\mathbb{E}_{\mathbf{H} \sim \mathcal{D}_{\text{gen}}} \|\mathbf{H} - \hat{\mathbf{H}}\|_2^2,
\end{equation}
where $\hat{\mathbf{H}} = f_{dec}(f_{enc}(\mathbf{H}))$.
\State Transmit the trained encoder $f_{enc}(\cdot)$ to UEs within this region.
\end{algorithmic}
\end{algorithm}

\subsection{Gossip-GAN For Catastrophic Forgetting Problem}

When a specific UE moves to a new environment, the DNN usually lacks generalization capability and performs poorly in the new channel distribution. DNN needs to be retrained for the ability to perform CSI feedback in the new scenario. However, when returning to the past scenarios, the neural network performs poorly. To avoid this catastrophic forgetting phenomenon, we can include data from the old scenarios when training in the new scenario, which can preserve the neural network's adaptability to past scenarios. However, continuously storing a large amount of data is not memory efficient. In this case, we further explore the capabilities of the proposed Gossip-GAN framework.

To effectively address the above-mentioned problem of catastrophic forgetting, 
DNN must take into account the past state of the channel in the face of the new channel distribution. It is worth noting that the proposed framework can capture the channel distribution of the encountered environment.
Let $G_t(\cdot)$ denote the generator obtained under the $t$-th scenario using the proposed framework. $G_t(\cdot)$ is capable of generating channel data that aligns with the distribution of the corresponding scenario. By saving the model $G_t(\cdot)$, we retain the memory of the encountered channel distributions.
Thus, as shown in Fig.~6, we propose a new memory updating strategy in which we can save the generator models in the previous environment.
 More concretely, when facing a new scenario, we combine the current scenario's Gossip-GAN trained generator $G_t(\cdot)$ and the past scenarios' generator models $G_{0,\ldots, t-1}(\cdot)$ to generate a mixed dataset $\tilde{\mathcal{H}}_{0,\ldots, t}$ for CSI feedback training: 
 \begin{equation} \label{new1}
\tilde{\mathcal{H}}_{0,\ldots, t}=G_{0,\ldots, t}(\boldsymbol{z}),  \;\;
 \boldsymbol{z}\sim N(0,\mathbf{I}).
\end{equation}
 The mixed dataset $\tilde{\mathcal{H}}_{0,\ldots, t}$ contains information about the past channel distributions, thereby avoiding the catastrophic forgetting problem.
This approach only requires storing a small number of model parameters, saving memory, thus enhancing the robustness of the model against variations in channel conditions. Meanwhile, there is no storage of any real data, addressing potential privacy issues for users.

 \begin{figure}[t!]
    \centering
    \includegraphics[width=8cm]{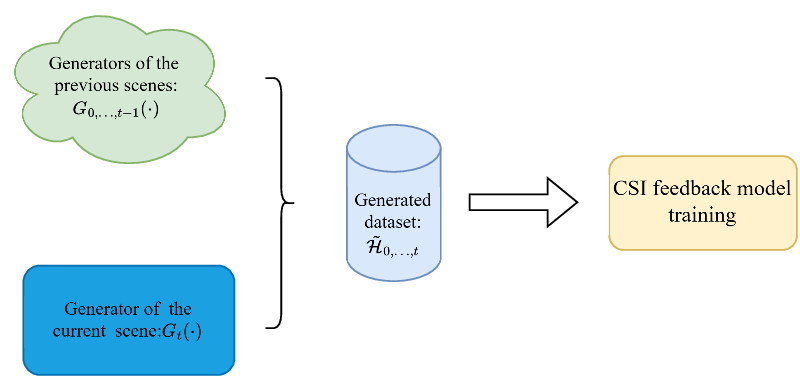}
    \caption{
    The schematic diagram of Gossip-GAN used to solve catastrophic forgetting.} 
    \label{fig:fig6}
\end{figure} 

\subsection{Scalability of the Proposed Gossip-GAN Training Framework}
Existing work \cite{ref16} reveals that deep learning for wireless communication typically involves two key steps: offline training and online deployment.  
In practical systems, as the UE density increases, it may even reach hundreds per cubic meter. Consequently, numerous UEs, such as IoT devices, may coexist within the same area (e.g., a factory or office) and share similar channel characteristics.  
During the offline training phase, the proposed Gossip-GAN framework selects a subset of $K$ UEs to train a CSI feedback model that achieves superior performance under the given channel characteristics. During online deployment, the trained \(f_{\text{enc}}(\cdot)\) is transmitted to numerous UEs across the region (as shown in Fig. 4), enabling broad application of the Gossp-GAN training framework.\footnote{When UE is a user and considering the UE’ mobility, the performance gain of the proposed Gossip-GAN framework against the DAE-based CSI feedback models still holds. In this case, the environment around the UE remains relatively stable as that around the BS. Note that the proposed framework is better suited for slow-moving scenarios, such as IoT devices gradually changing their locations between rooms in a factory during different working periods.} This demonstrates the scalability of the proposed framework.

The computational complexity of the proposed Gossip-GAN framework is incurred by two major operations performed in \textbf{Algorithm \ref{alg2}}, i.e.,  i) the GAN model training conducted in \textbf{Algorithm \ref{alg1}} and ii) the GL topology updating operation. Specifically, the computational complexity order of the GAN training is \(\mathcal{O}(N_{t}N_{c})\). The computational complexity order of the GL topology updating is \(\mathcal{O}(n_{\text{peers}}KT\log(N_{t}N_{c}))\) with \(T\) denoting the number of communication rounds per UE. Accordingly, the proposed Gossip-GAN framework requires an $
\mathcal{O}\left(N_{t}N_{c} + n_{\text{peers}}KT\log(N_{t}N_{c})\right)$
computational complexity.

\section{Simulation Results}

\begin{figure}[t!]
        \centering
        \includegraphics[width=8.5cm]{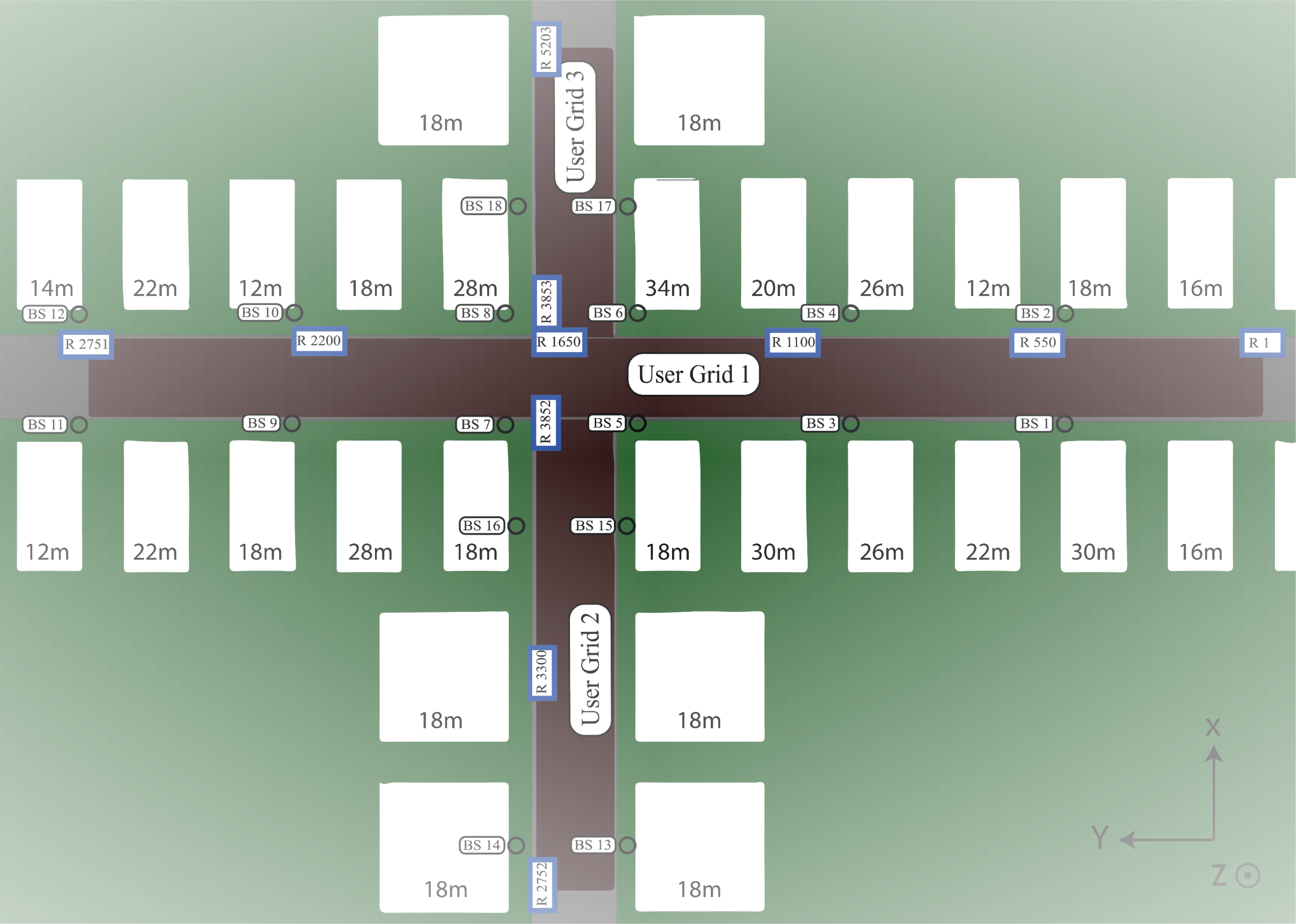}
        \caption{“O1\_28" an outdoor scenario of two streets and one intersection at operating frequencies 28 GHz
        in \textit{DeepMIMO}.}
        \label{fig:sub1}
    \label{fig:scene}
\end{figure}

\subsection{Simulation Specification}

In this section, we introduce hyperparameters and dataset generation settings in our experiments.

\begin{enumerate}
    \item \textit{Hyperparameter Settings}: 
     For the autoencoder model, we set the learning rate to be 0.001; the compression ratio $\gamma$ to be 1/16 by default; 
     and the training epoch to be 100. For the GAN model, we set the learning rate to be 0.001 and $\lambda_1=$ 10, $\lambda_2=$ 2, $M^\prime=$ 0.2, the dropout probability is 0.5, and the training epoch is 1000. The batchsize is 100. Besides, we use Adam as the optimizer.
    
    \item \textit{Dataset Settings}:
    \textit{DeepMIMO} is a publicly available dataset generated by the
Remcom Wireless Insite tool\cite{deepmimo}. As shown in Fig. \ref{fig:scene}, in our experiments, we separately verify the feasibility of the proposed framework in two scenarios:
\begin{itemize}
    \item \textit{Dense area}: The rows from 3540 to 3600, with a row spacing of 1 (0.2 m spacing),
corresponding to base station BS15 in the user grid2;
\item \textit{Sparse area}:
The rows from 1000 to 1400, with a row spacing of 6 (1.2 m spacing),
corresponding to base station BS4 in the user grid1.
\end{itemize}
In addition, we set the bandwidth to be 0.05 GHz, the transmitting antennas to be 32, the number of subcarries to be 32, and the number of
paths to be 25. 
Except for the special mentioned cases, we use the first 10000 CSI data in \textit{dense area} (\textit{sparse area}) as the dataset, with 5000 for training and 5000 for testing. Moreover, the data are normalized to the range of [-1,1]. 
\end{enumerate}

The normalized MSE (NMSE) is adopted as the metric:

\begin{equation}
\mathrm{NMSE}=\mathbb{E}\left[\frac{\|\widehat{\mathbf{H}}-\mathbf{H}\|_2^2}{\|\mathbf{H}\|_2^2}\right].\end{equation} 
\subsection{Performance of The Proposed Gossip-GAN Framework}

We first select the first 1000 CSI samples from the \textit{sparse area}, using 500 for training and 500 for testing. Similarly, we use 500 training and 500 testing samples from the widely used \textit{COST2100} dataset\cite{ref7} (i.e., the indoor picocellular scenario at the 5.3 GHz
band). Experiments are then conducted under different dropout probabilities to verify the validity of the adopted loss function. We use torchviz\cite{torchviz} to calculate the number of nodes in the model compute graph (the smaller the number, the lower the computational complexity). As shown in TABLE \ref{tab:sim1}, compared with CTGAN \cite{ref22}, the adopted loss functions (\ref{equ:ge}) and (\ref{equ:9}) can slightly reduce the size of the computation graph, while incurring only a small performance loss across different datasets and dropout probability settings.
\begin{table}[t!]
\centering
\caption{Performance comparisons of different loss functions using \textit{COST2100} dataset \cite{ref7} and  \textit{DeepMIMO} dataset \cite{deepmimo}.}
\resizebox{\columnwidth}{!}{
\begin{tabular}{|l|c|c|c|c|}
\hline
\textbf{Data source} & \textbf{Dropout probability} & \textbf{NMSE} & \textbf{Parameters G/D} & \textbf{Computation Graph Size} \\
\hline
True CSI of \textit{DeepMIMO} & /   & -11.59dB & 0.455M / 0.354M & / \\
\hline
\multirow{3}{*}{Fake CSI by (12) and (13) of \textit{DeepMIMO}} 
    & 0   & -9.24dB  & 0.455M / 0.354M & 1661 \\
\cline{2-5}
    & 0.3 & -10.63dB & 0.455M / 0.354M & 1661 \\
\cline{2-5}
    & 0.5 & -11.10dB & 0.455M / 0.354M & 1661 \\
\hline
\multirow{3}{*}{Fake CSI by (12) and (14) of \textit{DeepMIMO}} 
    & 0   & -9.36dB  & 0.455M / 0.355M & 1643 \\
\cline{2-5}
    & 0.3 & -10.47dB & 0.455M / 0.357M & 1643 \\
\cline{2-5}
    & 0.5 & -10.97dB & 0.455M / 0.357M & 1643 \\
\hline
True CSI of \textit{COST2100} & /   & -23.58dB & 0.455M / 0.354M & / \\
\hline
\multirow{3}{*}{Fake CSI by (12) and (13) of \textit{COST2100}} 
    & 0   & -18.53dB & 0.455M / 0.354M & 1661 \\
\cline{2-5}
    & 0.3 & -19.47dB & 0.455M / 0.354M & 1661 \\
\cline{2-5}
    & 0.5 & -20.14dB & 0.455M / 0.354M & 1661 \\
\hline
\multirow{3}{*}{Fake CSI by (12) and (14) of \textit{COST2100}} 
    & 0   & -18.17dB & 0.455M / 0.355M & 1643 \\
\cline{2-5}
    & 0.3 & -19.11dB & 0.455M / 0.356M & 1643 \\
\cline{2-5}
    & 0.5 & -19.53dB & 0.455M / 0.357M & 1643 \\
\hline
\end{tabular}
}
\label{tab:sim1}
\end{table}

Next, we verify the validity of the proposed framework. The proposed framework sets the number of users to 10, i.e., $K$ = 10, the number of data collected by each user to 500, and the two topologies shown in Section III are used to connect users every 10 epochs. We compare the NMSE performance of our proposed Gossip-GAN  framework with three counterparts: i) the GAN centrally trained using 5000 data, ii) the GAN distributively trained with 500 CSI per UE in a non-connected fashion, and iii) the CsiNet trained via FL. For the FL-based CsiNet, each UE communicates with the BS after the training of each epoch is completed and the CsiNet training epoch is also 100, to avoid high communication overhead. The resulting GAN model from these methods generates 5000 fake data for autoencoder training. 
\begin{figure}[t!]
    \centering
    \begin{subfigure}[b]{0.5\textwidth}
        \centering
        \includegraphics[width=\textwidth]{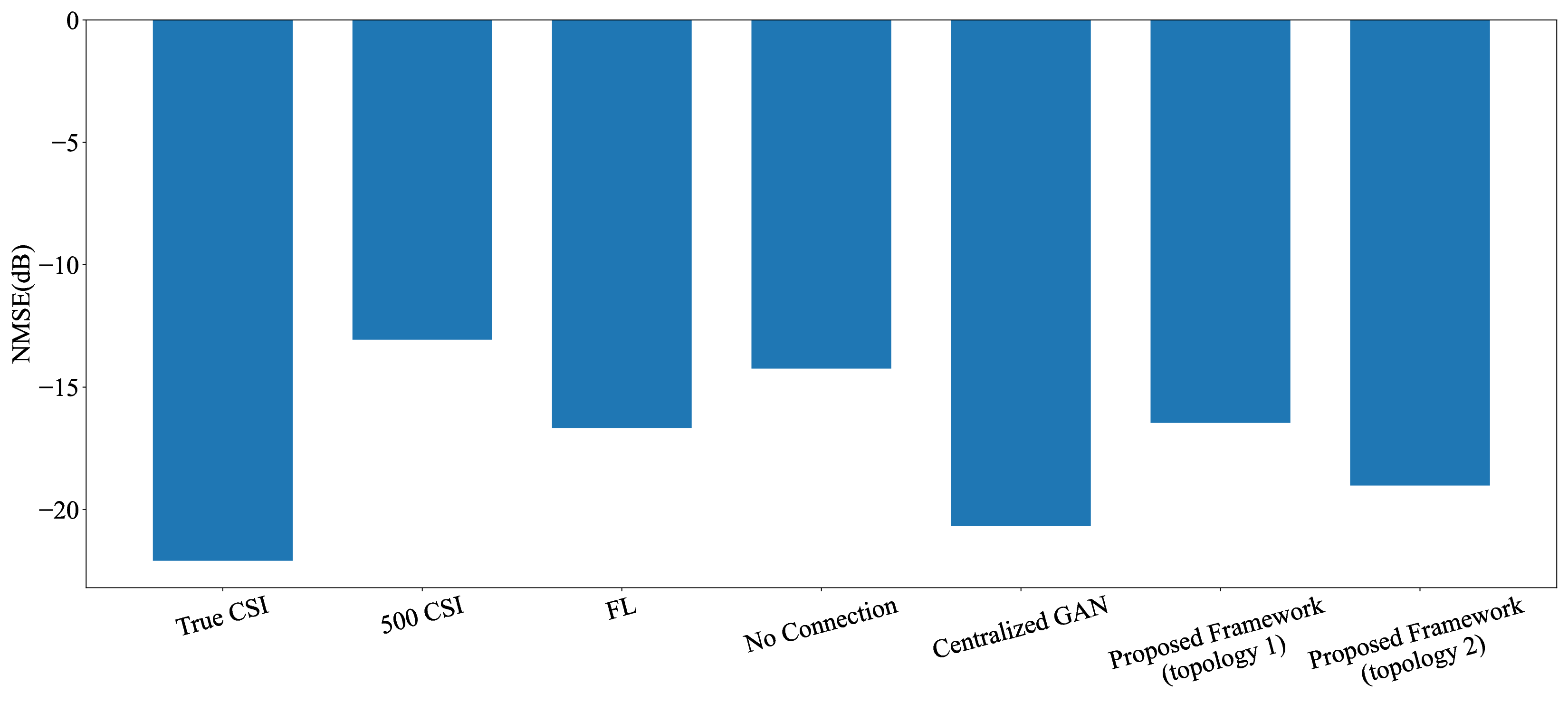}
        \caption{The NMSE (dB) performance in \textit{dense area}.
        }
        \label{fig:indoor}
    \end{subfigure}
    \quad
    \begin{subfigure}[b]{0.5\textwidth}
        \centering
    \includegraphics[width=\textwidth]{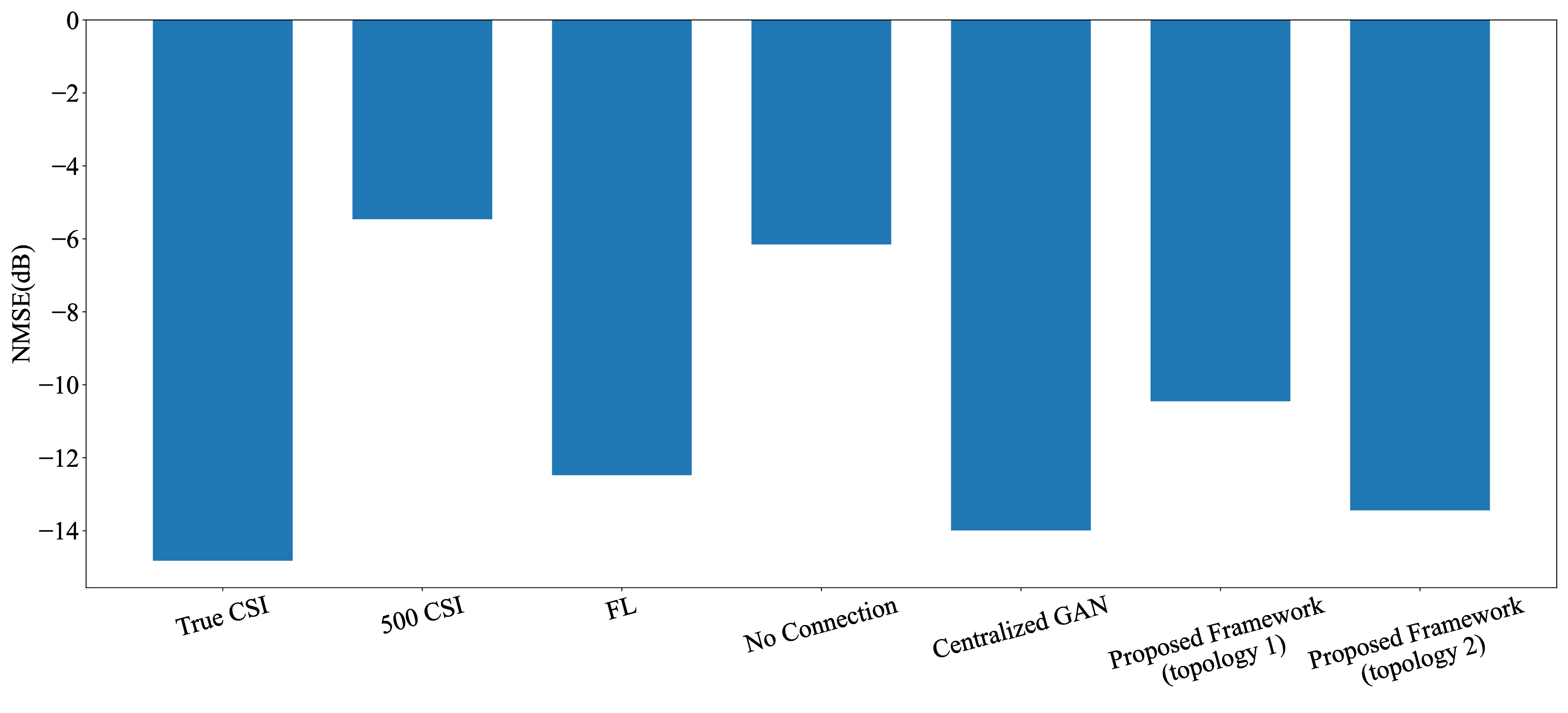}
        \caption{The NMSE (dB) performance in \textit{sparse area}.
        }
        \label{fig:outdoor}  
    \end{subfigure}
    \caption{The NMSE performance comparisons among different methods.}
    \label{fig:res1}
\end{figure}

\begin{table}[t!]
  \centering
\caption{Comparison of training costs between Gossip-GAN and centralized GAN, assuming each element is stored
as 4 bytes.}
  \resizebox{\columnwidth}{!}{
  \begin{tabular}{|c|c|c|c|}
    \hline
    \textbf{Method} & \textbf{FLOPS} & 
    \textbf{Memory Consumption} & \textbf{Per-Epoch Time Consumption} \\
    \hline
    Proposed Gossip-GAN & 0.98T & 7.142M & 2.977s \\
    \hline
    Centralized GAN & 9.8T & 42.296M & 25.784s \\
    \hline
  \end{tabular}}

  \label{tab:complexity_comparison}
\end{table}

As shown in Fig. \ref{fig:res1}, our proposed framework performs better than the case of no connection, especially in the case of sparse areas. In addition, we found out that the proposed framework using the second topology can approximate the performance of centralized GAN. However, comparing with the centralized training method, the proposed framework requires less CSI data collected by UEs, and the model training complexity is smaller.\footnote{
Fewer CSI are collected per UE, which means fewer channel estimation is performed, further reducing UE computation. Meanwhile, it is also worthwhile pointing out that when the number of UE increases, the reduction in computation will be more obvious.
} As shown in TABLE \ref{tab:complexity_comparison}, we compare the computational complexity, memory consumption, and the average execution time per epoch of the two approaches. The proposed framework adopts a distributed strategy, which effectively reduces computational complexity and memory usage while accelerating the training process. Therefore, the proposed framework is more suitable for UE with limited resources in practical scenarios.
Besides, the proposed framework requires transmitting only a small number of neural network parameters (0.455M), eliminating the need to upload the collected CSI data via the uplink channel. Notably, as the scale of the antenna array increases, the convolution-based GAN generator structure's advantage of low uplink
transmission
overhead will become even more significant, making it particularly well-suited for mMIMO-OFDM CSI feedback systems.


Next, we conduct 20 sets of repeated experiments to perform statistical validation on the proposed framework. 
Meanwhile, we present the 95\% confidence intervals of these two metrics, i.e., i) the NMSE (dB) and ii) the variance of Loss $\sigma^2$, in TABLE \ref{tab:confidence_interval_comparison}. In the repeated experiments, the convergence and performance of the proposed framework are akin to those achieved when using true CSI data. Meanwhile, the NMSE performance gap between the proposed framework and that obtained by training the autoencoder with real CSI is minimal.

\begin{table}[t!]
    \centering
        \caption{Statistical information of NMSE performance and $Loss$ variance $\sigma^2$.} 
    \resizebox{\columnwidth}{!}{
    \begin{tabular}{|l|l|l|l|}
        \hline
        \textbf{Method} & \textbf{Statistical Indicator} & \textbf{95\% Confidence Interval (Sparse Area)} & \textbf{95\% Confidence Interval (Dense Area)} \\
        \hline
        \multirow{2}{*}{\textbf{Proposed framework}} 
        & \textbf{NMSE (dB)} & ($-13.014$, $-12.697$) & ($-18.722$, $-18.419$) \\
        \cline{2-4} 
        & \textbf{Variance of Loss ($\sigma^2$)} & ($0.027$, $0.031$) & ($0.0093$, $0.0105$) \\
        \hline
        \multirow{2}{*}{\textbf{True CSI}} 
        & \textbf{NMSE (dB)} & ($-14.670$, $-14.338$) & ($-22.232$, $-21.791$) \\
        \cline{2-4} 
        & \textbf{Variance of Loss ($\sigma^2$)} & ($0.029$, $0.033$) & ($0.0099$, $0.0117$) \\
        \hline
    \end{tabular}}

    \label{tab:confidence_interval_comparison}
\end{table}

 \begin{figure}[t!]
    \centering
    \includegraphics[width=0.5\textwidth]{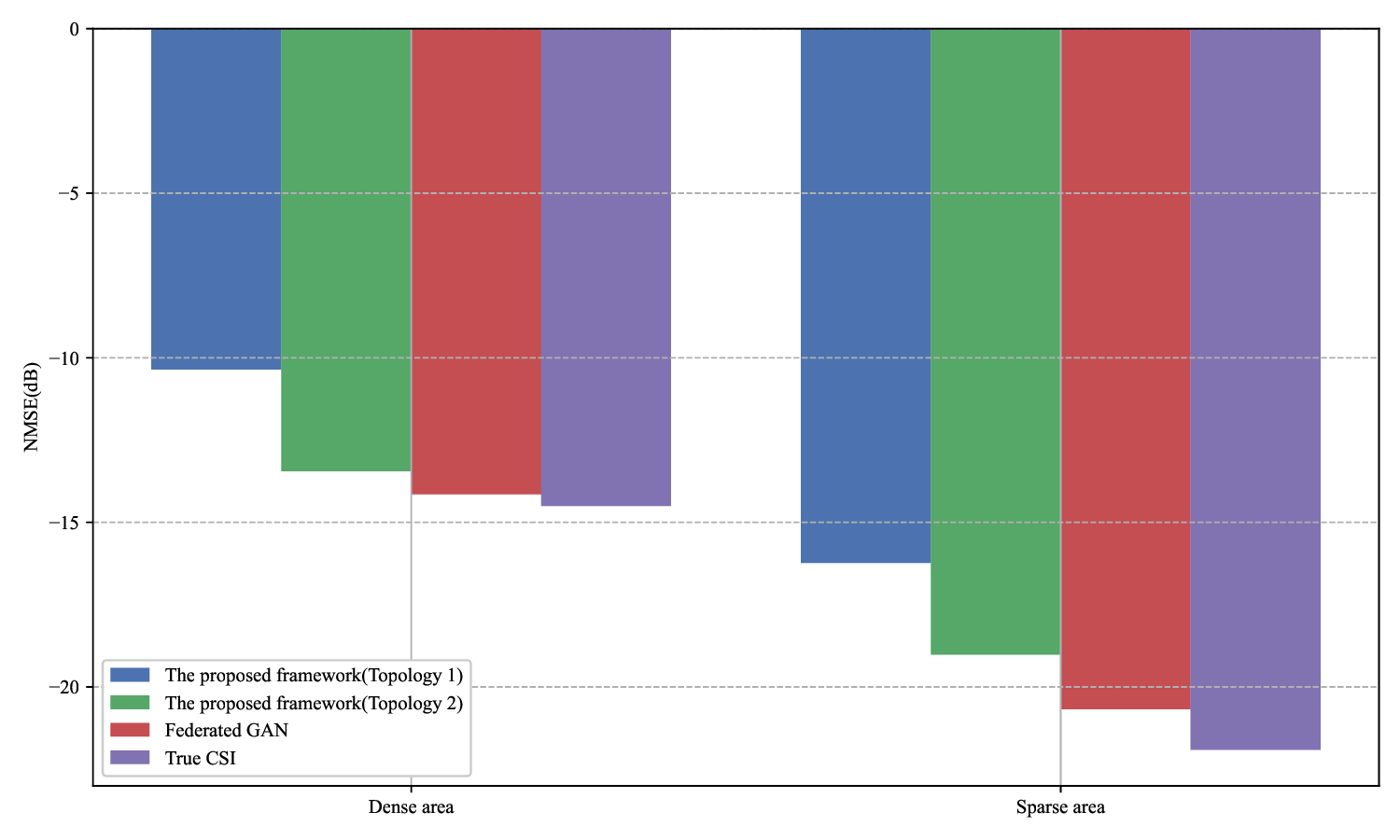}
    \caption{The NMSE performance comparison of the proposed framework with the federated GAN approach, i.e., \textit{Sync D\&G}\cite{federated-gan}.
    }
    \label{fig:figgangan}
\end{figure}

\begin{figure}[t!]
    \centering
    \begin{subfigure}[b]{0.5\textwidth}
        \centering
        \includegraphics[width=\textwidth]{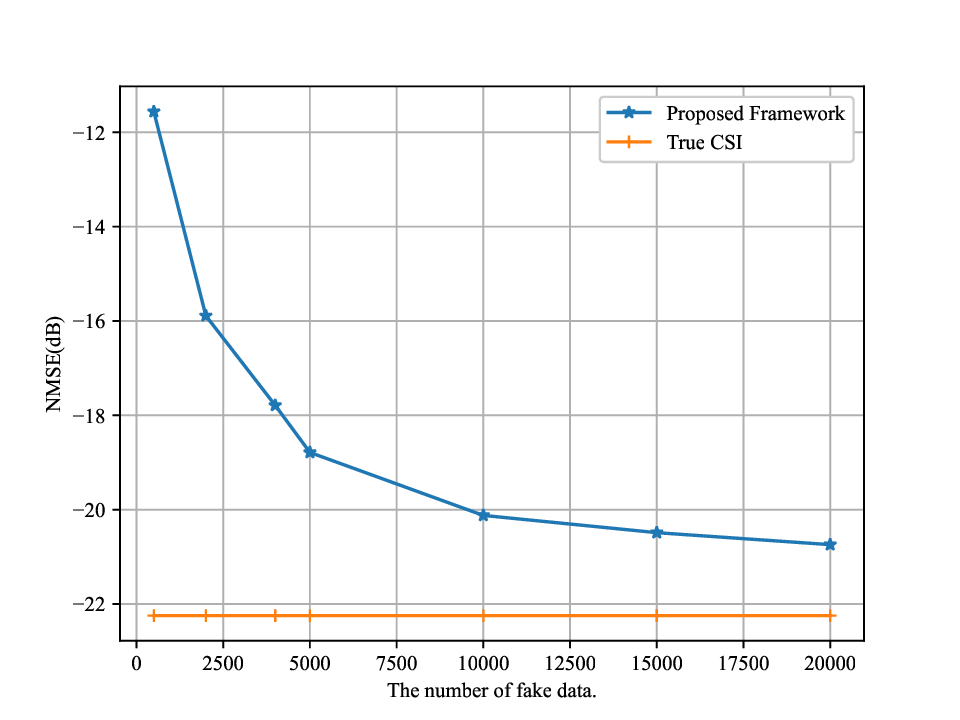}
        \caption{NMSE performance in the \textit{dense area}.}
        \label{fig:indoor12}
    \end{subfigure}
    \quad
    \begin{subfigure}[b]{0.5\textwidth}
        \centering
    \includegraphics[width=\textwidth]{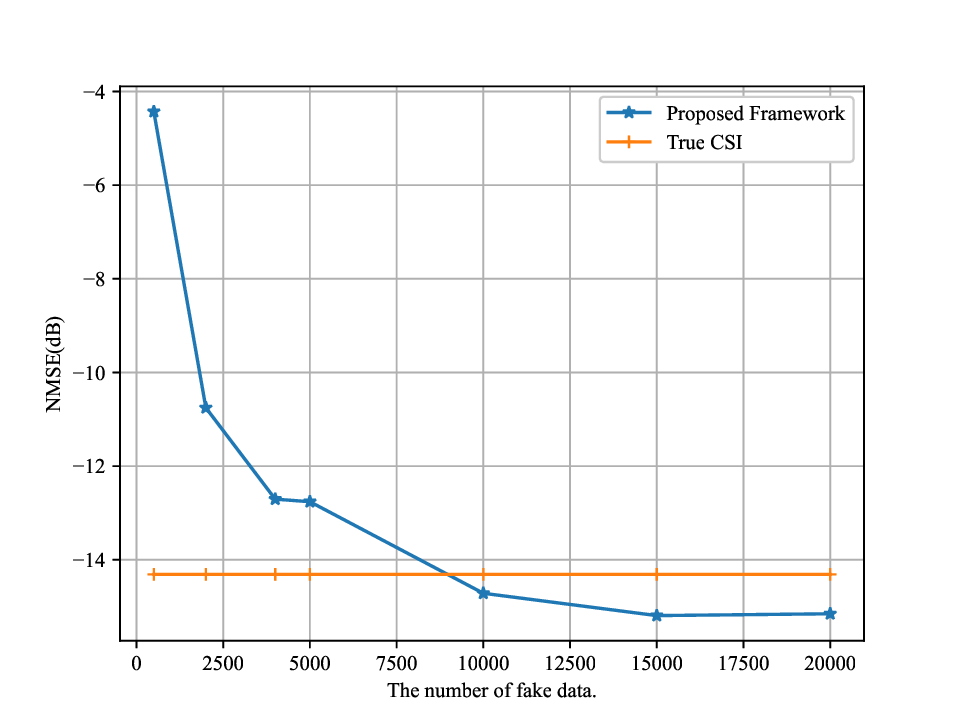}
        \caption{NMSE performance in the \textit{sparse area}.}
        \label{fig:outdoor12}
    \end{subfigure}
    \caption{NMSE performance vs. the number of fake data.}
    \label{fig:res3}
\end{figure}

In Fig. \ref{fig:res1} (a) and Fig. \ref{fig:res1} (b), NMSE with no connection GAN is better than NMSE with only 500 true CSI. This indicates that the number of fake data generated by GAN may improve the NMSE performance.
In Fig.~\ref{fig:figgangan}, we compare the NMSE performance of the proposed framework with that of the federated GAN, i.e., \textit{Sync D\&G}~\cite{federated-gan}. This figure reveals that the federated GAN achieves better NMSE performance than the proposed framework, as it utilizes the BS as a central server for the synchronizing and aggregating of models. However, this advantage comes at the cost of communication overhead with the BS, consuming excessive transmission bandwidth. Within each aggregation round, the federated GAN requires both uplink and downlink transmissions of the generator $G(\cdot)$ and discriminator $D(\cdot)$ models, with each communication involving 1.618M parameters.
In contrast, our framework adopts a fully decentralized training approach that eliminates the need for BS participation. It relies solely on communication among UEs,  thus significantly reducing bandwidth consumption at BS. 

Next, we use the proposed framework using the second topology (i.e., the fully connected mode) to verify this conclusion.

As shown in Fig. \ref{fig:res3}, we show the relationship between NMSE performance and the number of fake data in two areas. When the number of fake data is less than 1.0$e^{4}$, the performance of NMSE is significantly improved with the increase of the number of fake data, while when the number of fake data is larger than 1.0$e^{4}$, the performance of NMSE is not significantly improved and is gradually saturated.
An insufficient quantity of fake data may result in substantial NMSE performance degradation, whereas an excessive amount offers limited NMSE performance improvement with high training cost. To ensure high NMSE performance while avoiding the generation of excessive fake data, we set the amount of fake data to 1.0$e^{4}$ for practical efficiency.
In general, as the number of fake CSI increases, the performance of NMSE can be improved. Especially in sparse scenarios with limited data, the proposed framework can even slightly improve the performance of NMSE compared with true CSI. This proves that our proposed framework can model CSI that approximates the current channel distribution.

\begin{figure}[t!]
    \centering
    \begin{subfigure}[b]{0.5\textwidth}
        \centering
        \includegraphics[width=\textwidth]{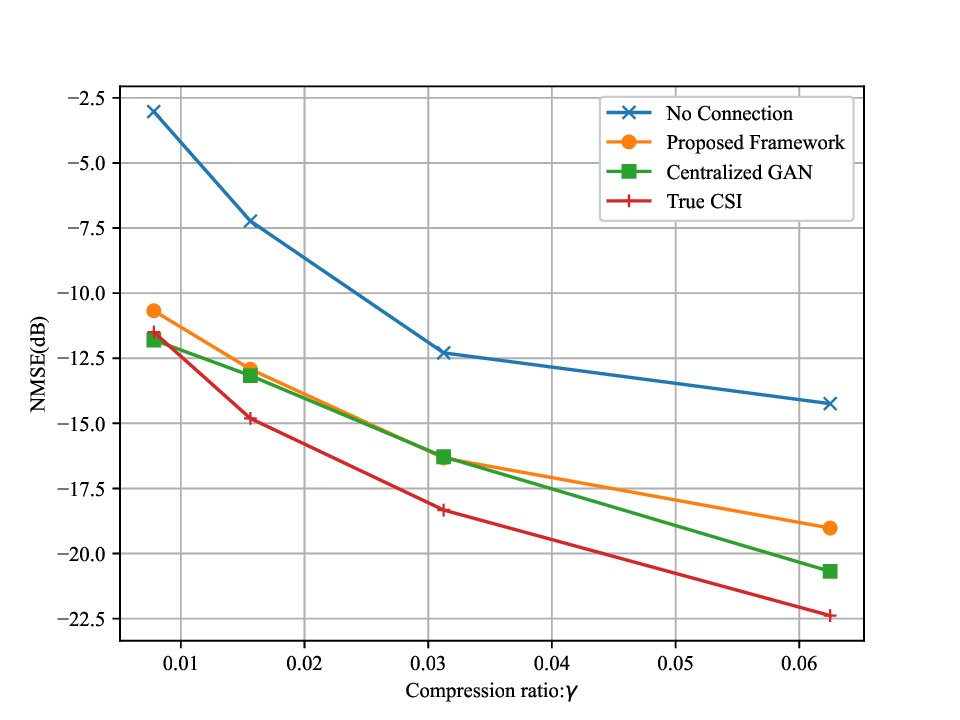}
        \caption{NMSE(dB) in the \textit{dense area}.}
        \label{fig:indoor12}
    \end{subfigure}
    \quad
    \begin{subfigure}[b]{0.5\textwidth}
        \centering
    \includegraphics[width=\textwidth]{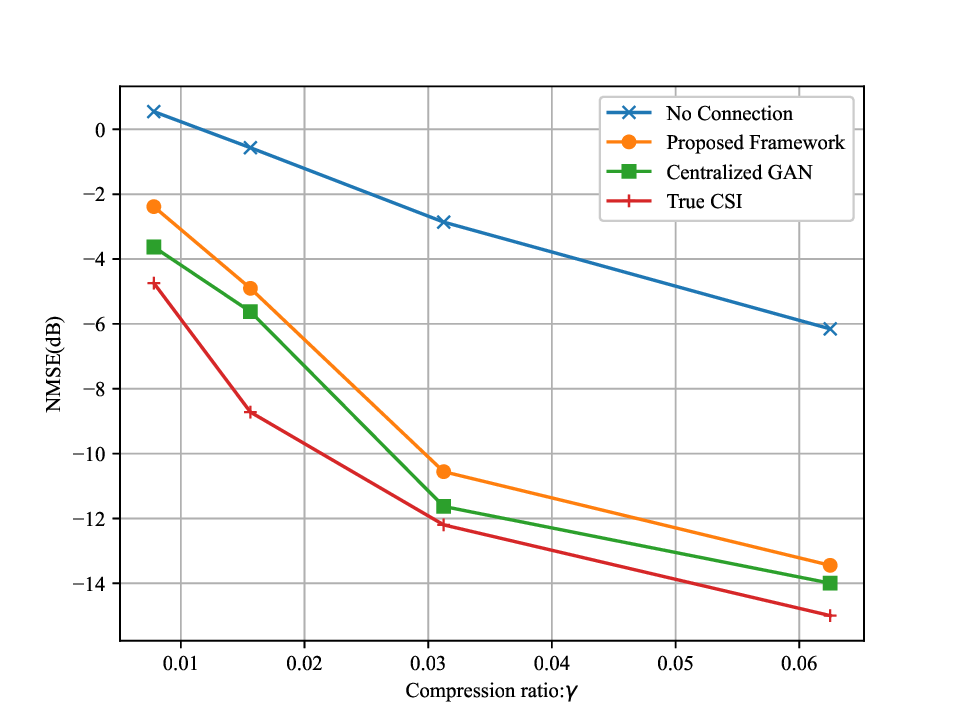}
        \caption{NMSE(dB) in the \textit{sparse area}.}
        \label{fig:outdoor12}
    \end{subfigure}
    \caption{NMSE performance against compression ratio $\gamma$.}
    \label{fig:res2}
\end{figure}

In the following, we set the compressed ratios $\gamma$ in (\ref{equ:5}) to be 1/16, 1/32, 1/64, and 1/128, respectively. Fig. \ref{fig:res2} shows how the NMSE performance varies with the compressed ratio $\gamma$ in two areas. The proposed framework adopting the second topology can use crowd intelligence, and its performance is always better than that of the unconnected method at different compressed ratios $\gamma$. The NMSE gap between the proposed framework, centralized training GAN, and true CSI under different environments and compression ratios $\gamma$ remains within an acceptable range. It demonstrates the robustness of our proposed framework.

\begin{figure}[t!]
    \centering
    \includegraphics[width=0.5\textwidth]{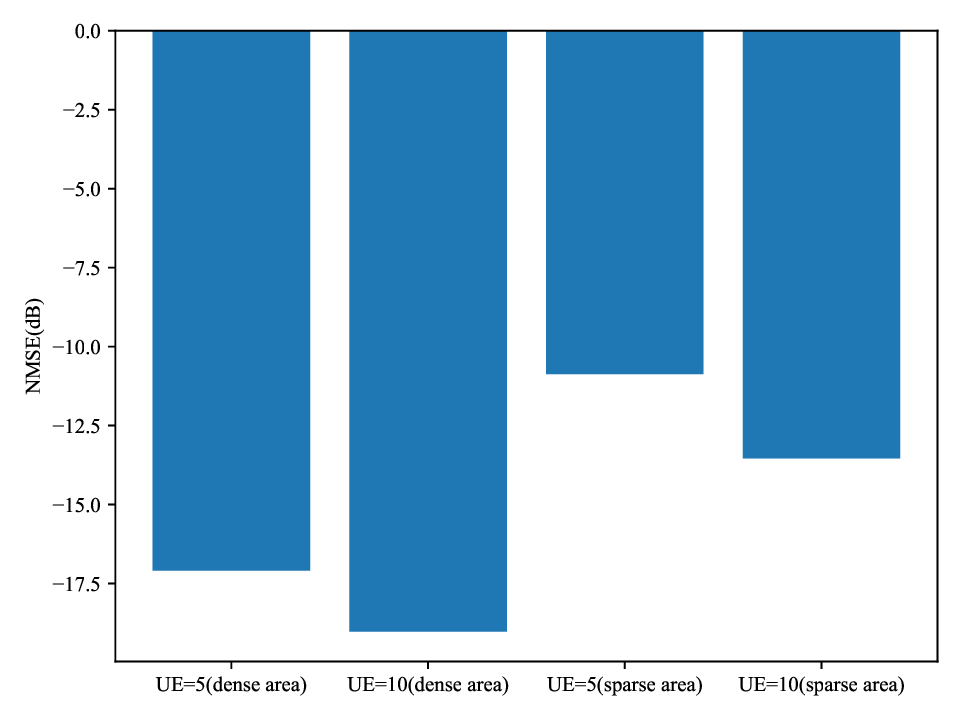}
    \caption{NMSE (dB) performance of the proposed framework in the case of different numbers of UEs.}
    \label{figueue}
\end{figure}

In the following, we verify 
the NMSE performance of our proposed framework impacted by the factor $K$, i.e.,
the number of UEs. Our proposed framework adopts the second topology (i.e., the fully connected mode), and each UE collects 500 local CSI in the specific area. 
The proposed framework aims to leverage crowd intelligence. When more UEs are involved during training, the Gossip-GAN can obtain a richer set of CSI through decentralized communication and model merging among UEs. As shown in Fig. \ref{figueue}, in both scenarios, as the number of UEs increases from 5 to 10, the NMSE obtained by the proposed framework also improves. The proposed framework can collect the model knowledge among different UEs by means of D2D, and finally obtain the approximate CSI distribution in the current environment. The more UEs, the more accurate CSI distribution can be modeled. However, involving more UEs will result in increased communication overhead.
Similar to federated learning, the proposed framework enables distributed model training. However, our proposed framework works without the involvement of a central processor (i.e., the BS)\cite{hegedHus2019gossip}, as a completely distributed approach that does not require frequent uplink switching to the BS.

 \begin{figure}[t!]
    \centering
    \includegraphics[width=0.5\textwidth]{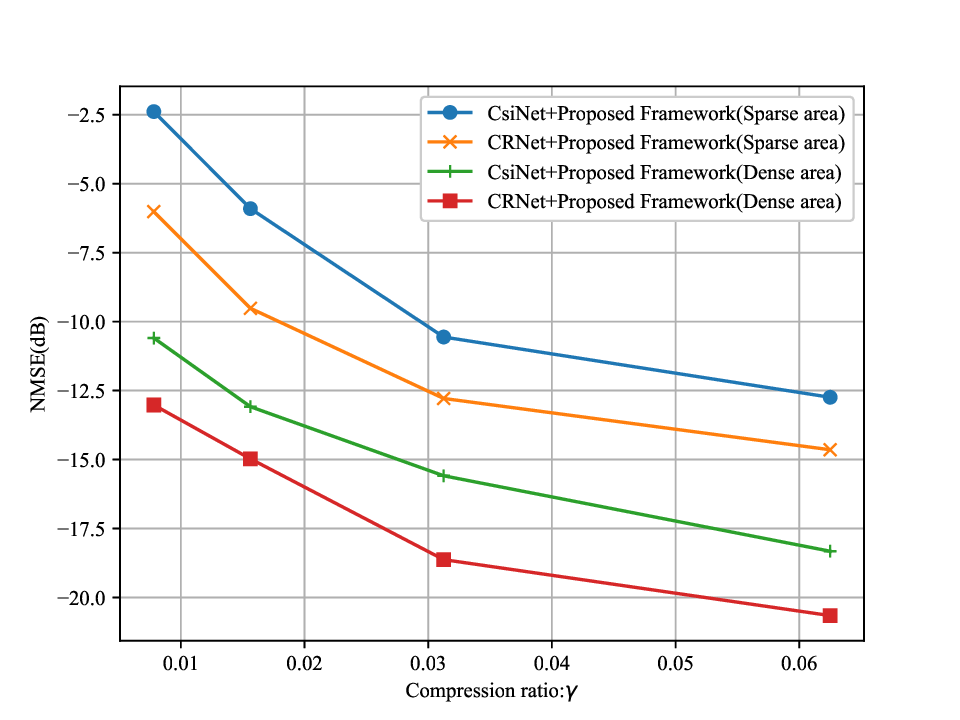}
    \caption{
Comparison of NMSE performance after coupling the proposed framework with other technologies.
    }
    \label{fig:fig711}
\end{figure}

In the above experiments, the autoencoder structure we adopted is the original CsiNet, while recently many new autoencoder techniques have been applied to CSI feedback problems and achieved excellent results. It is worthwhile noting that the proposed framework for CSI feedback training is compatible with other state-of-the-art technologies and is not limited to the CsiNet structure. Next, we combine the proposed framework with CRNet to conduct the experiments. Recall that
CRNet uses the cosine annealing learning rate as \cite{ref8}. The initial learning rate is set to be 0.1, and the number of epochs is set to be 100. As shown in Fig. \ref{fig:fig711}, our proposed framework combined with the CRNet further improves the NMSE performance compared to CsiNet.
The proposed framework can be seamlessly integrated with other advanced CSI feedback models, demonstrating the robustness of our framework.


 \begin{figure}[t!]
    \centering
    \includegraphics[width=0.5\textwidth]{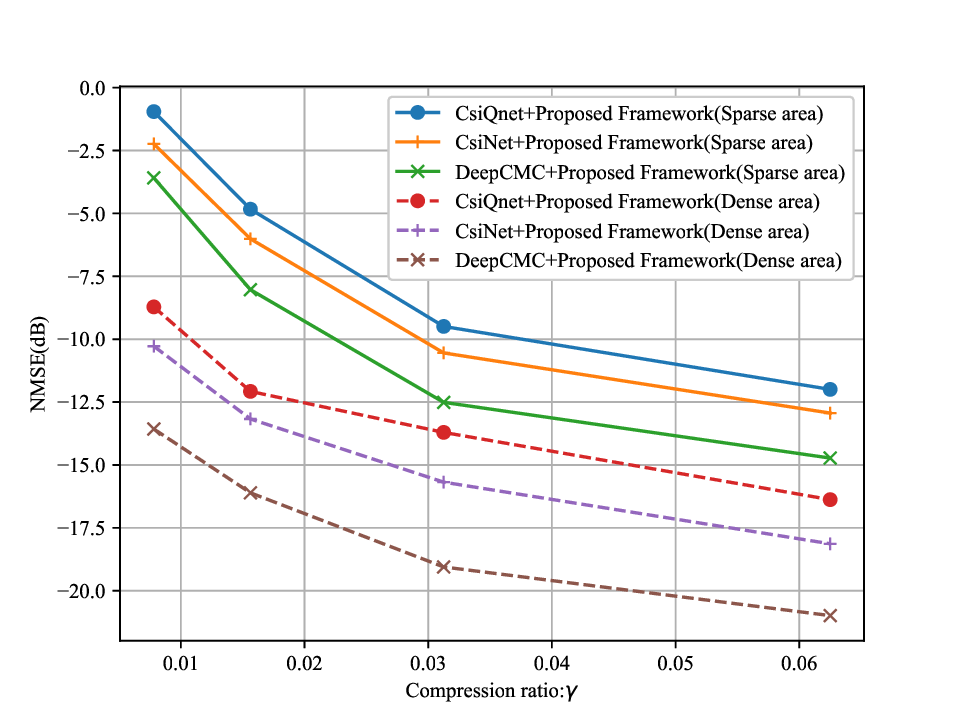}
    \caption{NMSE performance comparison of the proposed framework integrated with the state-of-the-art CsiQNet\cite{qnet}, CsiNet, and DeepCMC\cite{deepcmc} architectures, both of which undergo low-rate quantization. Specifically, during the model training, the $\lambda$ parameter of DeepCMC is set to $5\times10^{6}$, and CsiQNet is quantized to 5 bits.}
    \label{fig:fig712}
\end{figure}

In Fig. \ref{fig:fig712}, we compare the NMSE performance of the proposed framework integrated with the state-of-the-art CsiQnet \cite{qnet}, CsiNet, and deep learning-based CSI compression scheme (DeepCMC) \cite{deepcmc} architectures. As shown in Fig. \ref{fig:fig712}, the proposed Gossip-GAN framework for CSI feedback training is compatible with the state-of-the-art CsiQnet, CsiNet, and DeepCMC architectures. In addition, we observe from Fig. \ref{fig:fig712} that the proposed Gossip-GAN framework together with DeepCMC enables superior NMSE performance than Gossip-GAN with CsiQnet, CsiNet, thus leading to higher feedback accuracy performance.

Similar to most existing works (without retraining, as shown in TABLE \ref{tab:end}), CsiNet achieves great feedback performance when using the true CSI data collected from the \textit{sparse area}.
However, when the user moves to a new scene (\textit{dense area}), 
the model lacks generalization ability in new scenarios and performs poorly.
Thus, it is essential to train the model using newly collected CSI data to enable its adaptation to this context.
However, it is notable that DNN forgets the data characteristics of the previous scenario (sparse area) : The NMSE performance is only -0.33 dB. The feasible solution is to use the retraining method, when moving from \textit{sparse area} to \textit{dense area}, BS stores the true CSI data at \textit{sparse area}. Then, the CSI of \textit{dense area} and \textit{sparse area} are combined for training during \textit{dense area} training. As shown in TABLE \ref{tab:end}, this retraining method still achieves -15.89 dB NMSE performance in \textit{sparse area}. 
Then, in this work we store the generator model obtained in \textit{sparse area} in BS, and then when moving to \textit{dense area}, the BS combined the generator model previously stored with the current generator model obtained through the proposed framework and then each model generates a 1.0$e^{4}$ fake CSI dataset for CsiNet training. 
As shown in TABLE \ref{tablelast}, the proposed framework only needs to store generator model parameters, which is memory-efficient than directly storing the CSI data.
Through this method, the NMSE performance of the CSI feedback framework in the two areas is -14.75 dB and -18.51 dB respectively, which addresses the problem of catastrophic forgetting and achieves the feedback performance comparable to those of the retraining method.

Following the results in TABLE \ref{tab:end} and TABLE \ref{tablelast}, we found out that: i) the proposed Gossip-GAN framework can achieve much better feedback accuracy than the DNN-based scheme; ii) The retraining-based scheme enables higher feedback accuracy than our framework, as it uses the CSI collected from the new environment and the stored true CSI data from the past environment for training. However, the performance gain is achieved by sacrificing the computational cost and the memory cost. iii) We observe from TABLE \ref{tablelast} that, the proposed framework consumes lower memory cost than the retraining-based method.

\begin{table}[t!]
\centering
\caption{The NMSE performance when the UE moves from a \textit{sparse area} to a \textit{dense area}.}
\begin{tabular}{|c|c|c|c|}
\hline
\textbf{Method} & \textbf{After training on} & \multicolumn{2}{c|}{\textbf{NMSE (dB)}} \\ \cline{3-4} 
 &  & Sparse area & Dense area \\ \hline
\multirow{2}{*}{No retraining} & Sparse area & -15.04 & -0.54 \\ \cline{2-4} 
 & Dense area & -0.33 & -22.56 \\ \hline
\multirow{2}{*}{Retraining} & Sparse area & -15.04 & -0.54 \\ \cline{2-4} 
 & Dense area & -15.89 & -20.28 \\ \hline
\multirow{2}{*}{Proposed framework} & Sparse area & -14.84 & -1.29 \\ \cline{2-4} 
 & Dense area & -14.75 & -18.51 \\ \hline
 
\end{tabular}

\label{tab:end}
\end{table}

\begin{table}[t!]
\centering
\caption{The memory storage size in the current continual learning experiment is presented, assuming each element is stored as 4 bytes.}
\begin{tabular}{|c|c|c|}
\hline \textbf{Method} & Proposed framework & Retraining  \\
\hline \textbf{Memory cost} & 1.82 M & 39.06 M \\
\hline
\end{tabular}
\label{tablelast}
\end{table}

\section{Conclusions and Future Research Directions}
In this work, we proposed a fully distributed Gossip-GAN framework for CSI feedback training in FDD mMIMO-OFDM systems. The proposed framework utilizes crowd intelligence to collaboratively obtain a GAN model that can capture the context channel distribution, and then transmit the GAN's generator to the BS to generate a synthetic dataset for CSI feedback training. The proposed framework does not require the uplink transmission of huge CSI data or the involvement of BS as a central server, solving the problem of uplink bandwidth consumption and potential user privacy issues. Meanwhile, we proposed a fully distributed approach to reduce the computational complexity of GAN training. 
In the simulation section, we adopted the \textit{DeepMIMO}: “O1\_28” scenario, a dataset in the millimeter wave band, and conducted experiments to validate the feasibility of our proposed framework.
Simulation results demonstrated that the proposed framework is capable of addressing the issue of catastrophic forgetting for CSI feedback scenarios.

Based on this work, there are still some research directions that need to be further explored:

\begin{itemize}
\item 
It is possible to apply the proposed framework to the physical layer, such as the research areas in terms of the beamformer design, channel estimation, and data detection.
\item 
We will investigate advanced generative models to produce higher-quality synthetic CSI datasets.
\item 
In addition, it is worthwhile exploring an enhanced GL topology to reduce communication between D2D without compromising the system performance. Note that how to design advanced network connectivity topology taking into account the NMSE performance, the dynamic user environment, communication constraints, and synchronization issues in real-world heterogeneous networks will also be studied in our next-step research. 
Inspired by the \textit{Sync D\&G} federated learning framework \cite{federated-gan}, one promising approach we envision is to design a new algorithm for GL topology, in which a specific UE can be selected as the central node to coordinate synchronous model updates and training.
\item 
For the proposed framework, the significant risk posed by adversarial machine learning (ML) threats deserves to be addressed, especially given that GANs are known to be vulnerable to adversarial attacks.
In our next-step research, we will study how to ensure Gossip-GAN security against adversarial attacks in CSI feedback by adopting differential privacy, or multimodal defense strategies.
\item 
In our future research direction, we will adopt the 3GPP urban macro non-line-of-sight (NLOS) channel model \cite{docomo20165g} and then use the generated channel dataset under this channel model to further validate the robustness of the proposed Gossip-GAN framework.
\end{itemize}

\bibliographystyle{IEEEtran}%
\bibliography{bibfile}
\end{document}